\documentclass[%
 reprint,
 amsmath,amssymb,
 aps,
]{revtex4-2}

\usepackage{graphicx}% Include figure files
\usepackage{dcolumn}% Align table columns on decimal point
\usepackage{bm}% bold math
\usepackage{xcolor}
\usepackage{footmisc}
\newcommand{\dt}[0]{\bm{\Delta \theta}}
\newcommand{\dth}[0]{\bm{\Delta\hat{\theta}}}
\newcommand{\dy}[0]{\bm{\Delta y}}

\newcommand{\nsp}[0]{n_\mathrm{sp}}
\DeclareMathOperator{\rank}{rank}

\begin{document}

\preprint{APS/123-QED}

\title{Fully independent response in disordered solids}% Force line breaks with \\
%\thanks{Self-assembly of disordered solids}%

\author{Mengjie Zu}
% \altaffiliation{Institute of Science and Technology Austria}%Lines break automatically or can be forced with \\
 \email{mengjie.zu@ist.ac.at}
\author{Aayush Desai}
\author{Carl P. Goodrich}%
 \email{carl.goodrich@ist.ac.at}
\affiliation{%
 Institute of Science and Technology Austria (ISTA), Am Campus 1, 3400 Klosterneuburg, Austria
}%

\date{\today}% It is always \today, today,

\begin{abstract}
Unlike in crystals, it is difficult to trace emergent material properties of amorphous solids to their underlying structure. Nevertheless, one can tune features of a disordered spring network, ranging from bulk elastic constants to specific allosteric responses, through highly precise alterations of the structure. This has been understood through the notion of independent bond-level response -- the observation that in many cases, different springs have different effects on different properties. While this idea has motivated inverse design in numerous contexts, it has not been formalized and quantified in a general context that not just informs but enables and predicts inverse design. 
Here, we show how to quantify independent response by linearizing the simultaneous change in multiple emergent features, and introduce the much stronger notion of fully independent response. Remarkably, we find that the mechanical properties of disordered solids are always fully independent across a wide array of scenarios,
regardless of the target features, tunable parameters, system size, dimensionality, and class of interactions.
Furthermore, our formulation quantifies the susceptibility of features to parameter changes, which is correlated with the maximum linear tunability. We also demonstrate the implications for multi-feature inverse design beyond the linear regime. 
These results formalize our understanding of a key fundamental difference between ordered and disordered solids while also creating a practical tool to both understand and perform inverse design. 

\begin{description}
\item[keywords]
 disordered solids $|$ independent response $|$ inverse design $|$ elastic moduli $|$ automatic differentiation
\end{description}
\end{abstract}

\maketitle

Remove a single spring from a disordered spring network, and the bulk and shear moduli will change in different and uncorrelated ways~\cite{Goodrich2015,Hexner2017}. 
This notion, that different observables in a disordered solid have independent responses when altering the system at the bond level, has profound consequences as it implies a roadmap for inverse design where different observables are controlled separately. Indeed, this principle has motivated a wide range of work. For example, tuning the bulk and shear moduli separately has enabled the design and creation of auxetic materials~\cite{Goodrich2015, Hexner2017, Reid2018,Shen2021, Zaiser2023}, and microscopic responses have been targeted to create nontrivial stress patterns~\cite{Hexner2023} and allostery-like functionality in both mechanical and flow networks~\cite{Rocks2017,Rocks2019,Pashine2021,Pashine2023}. Furthermore, physical learning and other adaptive training protocols exploit the large design space enabled by independent response~\cite{Stern2021,Hagh2022,Falk2023,Stern2023,Mendels2023}. Finally, independent response at the species level has been shown to survive ensemble averaging, leading to the robust inverse-design of self-assembling disordered solids with precisely tuned properties~\cite{Zu2024}. 

Independent response in disordered solids clearly suggests that these materials are highly tuneable, but key critical questions remain. 
For example, it has been shown that the bulk and shear moduli can be tuned relative to each other, but the mechanics of finite-size disordered systems are actually described by 21, not two, elastic constants in three dimensions, as well as six components of the residual stress tensor, and order-$N$ modes of vibration, all of which one might want to control. Is there a limit to how many of these properties can be tuned simultaneously? Are some properties easier to tune than others, and how does this depend on the properties in question and the parameters being adjusted? 

While these questions have been addressed in specific contexts, we lack a robust and concrete methodology for calculating and quantifying independence that can address these questions in any context while also serving as a practical tool for inverse design.

In this letter, we demonstrate the utility of a straightforward formulation of independent response by means of the Jacobian, $J$, that relates changes in emergent features (e.g. elastic constants) to changes in model parameters (e.g. spring constants, particle diameters). 
Independence is quantified by the rank of $J$, which directly measures the number of features that can be tuned independently by adjusting the parameters. Surprisingly, we find that mechanical properties of disordered solids are ``fully independent'' in a wide variety of settings and contexts, meaning that any and all properties of interest can be tuned independently provided there are at least as many parameters. This does not have to be so, and is decidedly not the case for crystals, emphasizing how disorder interacts with elasticity.
Further insight can be gained from the singular value decomposition of $J$: the singular values measure the susceptibility of certain feature combinations to particular parameter combinations, providing direct and practical insight into the mechanics of tuning and inverse design. Thus, by focusing on the Jacobian, we provide a well-defined, rigorous, broadly applicable, and useful formulation of independent response, and reveal its far reaching consequences in disordered solids.

Rather than focusing only on spring networks, we consider a broad class of athermal particle packings that allows us to switch between spring networks and mechanically stable jammed packings with or without attractions. Furthermore, we parameterize the interactions at the species level so that we can seamlessly vary the number and type of tuneable parameters. More precisely, we mainly consider systems of $N$ particles in 2 and 3 dimensions with interactions given by 

\begin{align} \label{eq:hmorse}
    u_{ij}(r) = \begin{cases}
        \frac{k_{ij}}{2}(1 - r/\sigma_{ij})^2 - B_{ij}, & r < \sigma_{ij} \\
        B_{ij}\left( 1 - e^{-a(r - \sigma_{ij})} \right)^2 - B_{ij}, & r \geq \sigma_{ij},
    \end{cases}
\end{align}
where $r$ is the center-center distance between particles $i$ and $j$, $\sigma_{ij}=(D_i+D_j)/2$ is the mean of the two particle diameters, and $k_{ij}$ is a repulsive energy scale. We will often set $B_{ij}=0$, in which case this is the finite-ranged, purely repulsive ``soft-sphere" potential common in the study of jamming. When $B_{ij}\neq 0$, there is an attractive force given by the Morse potential whose magnitude and range are given by $B_{ij}$ and $a$, respectively. The parameters are determined at the species level, so a system with $\nsp$ species will be controlled by $\nsp$ particle diameters $D_i$, as well as $\nsp(\nsp+1)/2$ repulsive and attractive energy scales, $k_{ij}$ and $B_{ij}$, respectively.

Initial systems are prepared with $N$ particles evenly divided into $\nsp=2$ species with a 1:1.4 size ratio to avoid crystallization. 
Mechanically stable zero-temperature configurations are generated by placing particles randomly in a periodic simulation box and minimizing the energy to a local energy minimum using the FIRE algorithm~\cite{Bitzek2006,Bitzek2020}. 

We then calculate a list of observables, or features. Features we consider include the 6 (21) independent elastic constants in 2 (3) dimensions, the 3 (6) independent elements of the stress tensor, and the $\mathcal{O}(N)$ non-zero vibrational mode frequencies. Discussion of these calculations appears in the Supplementary Material~\footnote{See Supplemental Material at [url], which includes Refs.~\cite{SI1,SI2} \label{SI}
.}. We then choose $n_y$ of these features, which we combine into a ``feature vector'' $\bm y$. 

Our goal is to systematically probe how $\bm y$ can be adjusted by dynamically tuning the interaction parameters. 
Importantly, there are many ways in which one can change the initial parameters to affect the features. Despite preparing systems with only two species -- large and small -- we now randomly assign particles to new species types, ensuring that new species are comprised of only big or small particles.  
This species reassignment does not in any way affect the initial system, but it does change the parameterization: there are now $\nsp$ species-level diameters that can be changed independently, as well as $\mathcal{O}(\nsp^2)$ interaction energies. This approach will allow us to systematically control the number and type of relevant adjustable parameters. Going forward, we will choose many different such parameterizations, and combine $n_\theta$ of these species-level parameters into a ``relevant parameter vector" $\bm \theta$.

We are now in a position to study how the feature vector $\bm y$ changes when we perturb the relevant parameters $\bm \theta$. Expanding around
$\bm \theta_0$, the set of parameters at which the system was prepared, we have
\begin{align} \label{eq:linear_approx}
    \bm y = \bm y_0 + J \bm{\Delta \theta} + \mathcal{O}(\bm{\Delta \theta}^2),
\end{align}
where $\dt = \bm \theta - \bm \theta_0$ and $J$ is the $n_y \times n_\theta$ Jacobian matrix. 
To accurately calculate $J$, we use Automatic Differentiation (AD) techniques~\cite{Baydin2018,Rumelhart1986,Wengert1964} to propagate derivatives through the entire calculation of the features, including energy minimization. See SI for more details. 

\begin{figure}
    \centering
    \includegraphics[width=1.0\linewidth]{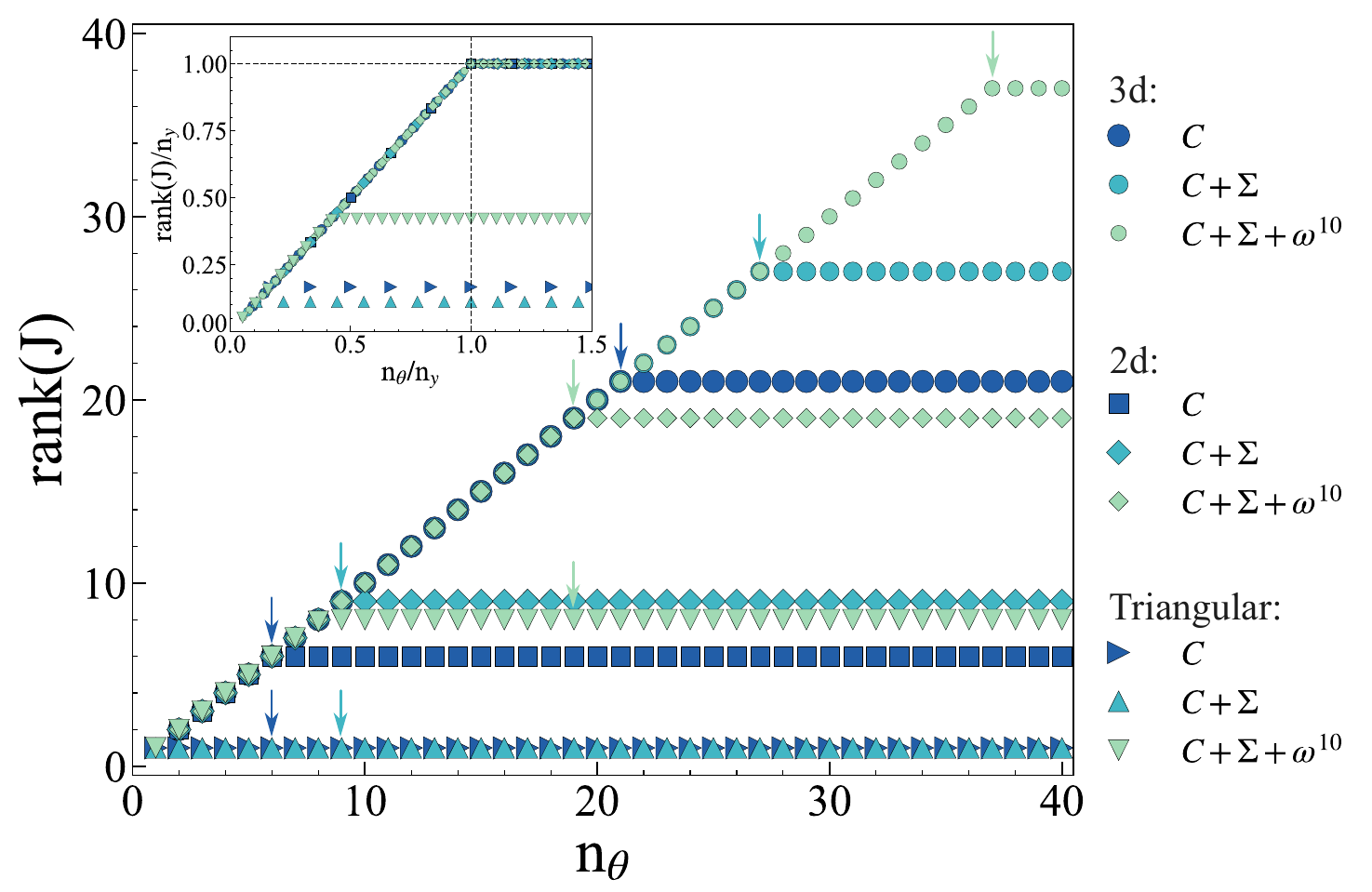}
    \caption{
    Fully independent response is revealed by $\rank(J)$, which is shown as a function of the number of parameters, $n_\theta$ for 9 different scenarios. The relevant features consist of different combinations of the elastic constants $C$, the elements of the stress tensor $\Sigma$, and the first 10 nonzero vibrational mode frequencies $\omega^{10}$. Circles and diamonds show data for 3d and 2d disordered packings composed of $N=256$ soft spheres ($B_{ij}=0$) with $k_{ij}=1$, and $p=10^{-1}$, while the triangles show data for crystals (2d, particles on a triangular lattice).     
    Particles are divided into $n_\theta$ species, as described in the text, the diameters of which are taken as the relevant parameters $\bm \theta$. 
    The vertical arrows point to the data points where $ n_\theta = n_y$.
    For all disordered systems, this arrow marks a cusp in behavior: $\rank(J)=n_\theta$ for $n_\theta \leq n_y$, and $\rank(J)=n_y$ for $n_\theta > n_y$. Thus, the $n_y$ features are always fully independent with respect to the $n_\theta$ parameters. 
    For crystals, $\rank(J)$ reaches an upper bound well before $n_\theta=n_y$. This stark difference is more clearly shown in the insert, where $\rank(J)$ and $n_\theta$ are scaled by $n_y$.
    }
    \label{fig:rankJ}
\end{figure}

\begin{figure}
    \centering
    \includegraphics[width=1.05\linewidth]
    {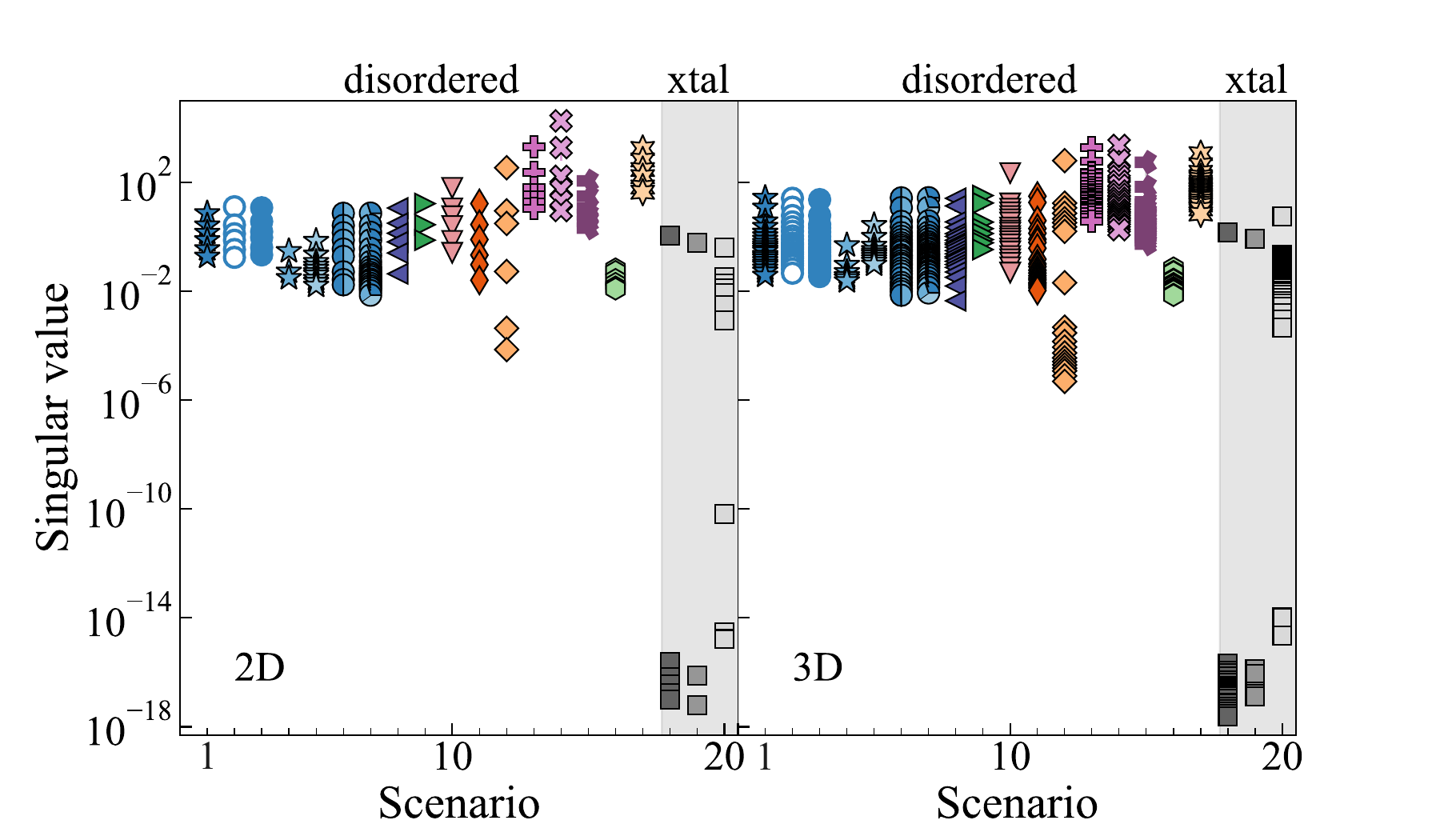}
    \includegraphics[width=1.0\linewidth]{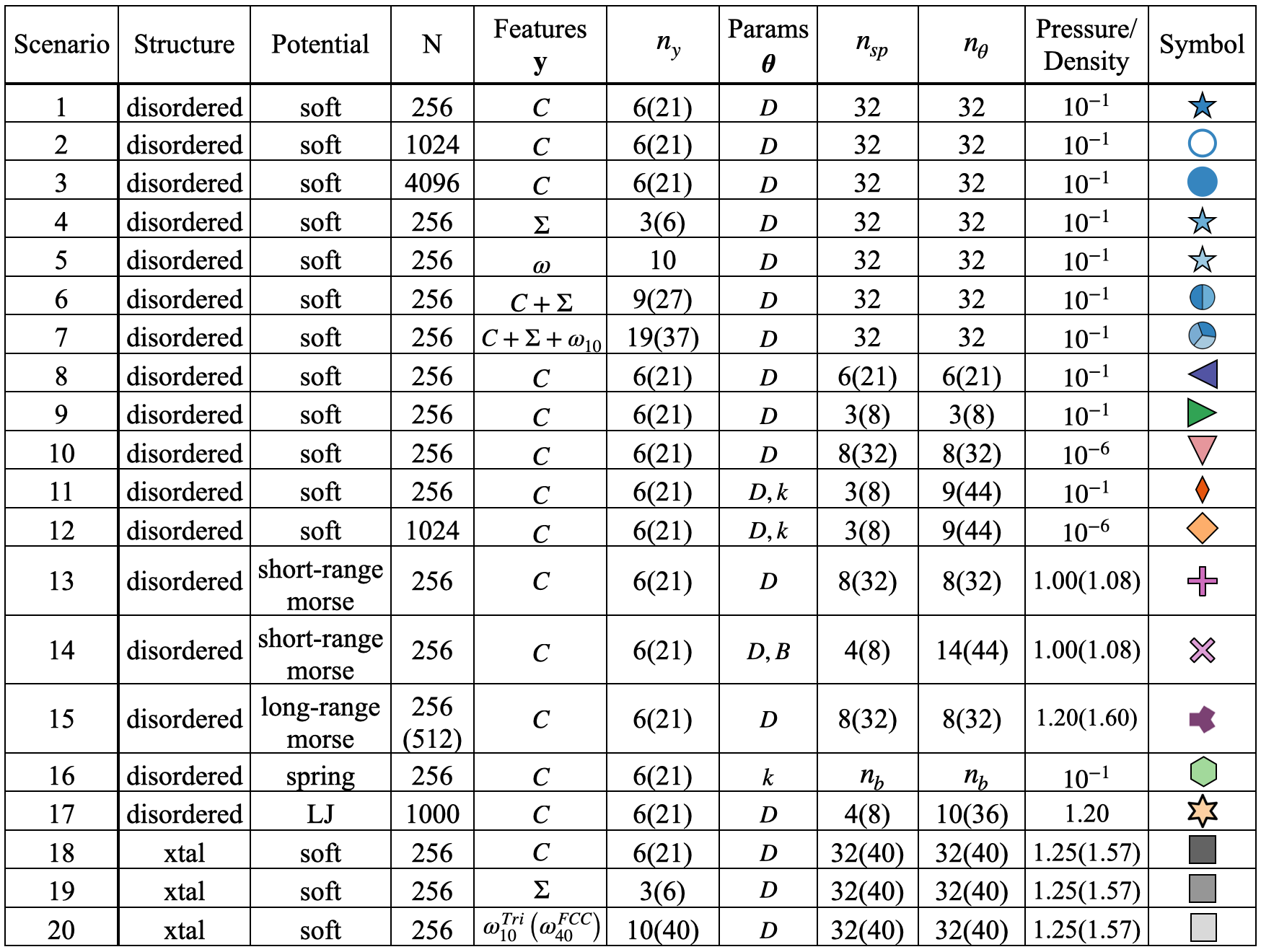}
    \caption{Susceptibilities, as measured by the singular values of $J$, reveal a qualitative difference between crystals and disordered solids. The plot shows all susceptibilities for 40 different scenarios with different features, parameters, potentials, system sizes, pressures, and structures. 
    For disordered systems, these susceptibilities are strictly positive, though they can vary by orders of magnitude, indicating that the features are fully independent. Data is averaged over 10 different systems for each scenario, and we have verified that all systems are fully independent individually. For crystals, however, only some of the singular values are positive, meaning the features are only partially independent.  The table describes the specifics of each scenario. Each scenario is repeated in $2d$ and $3d$, with identical setups except for when two numbers are shown, in which case they correspond to $2d$ ($3d$). A detailed description of each scenario is found in the Supplementary Material~\cite{Note1}.
    }
    \label{fig:singular_values}
\end{figure}

This Jacobian contains all the information necessary to quantify independent response, which is revealed through the singular value decomposition
\begin{align}
    J = U S W^T, 
\end{align}	
where $U$ is a $n_y$ by $n_y$ unitary matrix, $W$ is a $n_\theta$ by $n_\theta$ unitary matrix, and $S$ is a $n_y \times n_\theta$ matrix with nonzero elements only on the diagonal. The diagonal elements of $S$ are called the singular values and the columns of $U$ and $W$ are called the left and right singular vectors. The rank of $J$ is equal to the number of non-zero singular values and indicates the number of independent ways that the features can be altered through small changes to the parameters. More specifically, a small change in parameters $\dt$ in the direction of the $i$th right singular vector $w_i$ will lead to a change in features $\dy=\bm y - \bm y_0$ in the direction of the $i$th left singular vector $u_i$ scaled by the singular value $s_i$. Thus, the singular value $s_i$ can be interpreted as the susceptibility of the feature strain $u_i$ to design mode $w_i$. When $s_i$ is positive, we call the corresponding left and right singular vectors ``sets of compatible feature strains'' and ``relevant design modes,'' respectively, in part due to a strong connection with rigidity theory discussed in the Supplementary Material~\cite{Note1, Pellegrino1993}. When left and right singular vectors are not associated with a positive singular value, we call them ``sets of incompatible feature strains'' and ``irrelevant design modes.''

Calculating $\rank(J)$ quantifies the linear independence of the features $\bm y$ with respect to changes in the parameters $\bm \theta$. If $\rank(J) = n_y$, then all feature strains are compatible -- features can be tuned locally in any possible direction. 
Similarly, if $\rank(J) = n_\theta$, then all design modes are relevant and affect the features. Since the rank of a matrix can never be larger than the smaller dimension, $\rank(J) \leq \min(n_y,n_\theta)$. If  $\rank(J) = \min(n_y,n_\theta)$, then $J$ is a full rank matrix and we call the features ``fully independent'' with respect to the given parameters. If $\rank(J) < \min(n_y,n_\theta)$, then $\bm y$ is only ``partially independent,'' with independence given by the sets of compatible feature strains.

The primary result of this paper is that the mechanical properties of disordered solids, unlike those of crystals, are fully independent in a wide range of situations, robust to changes in the number and type of features, the number and type of relevant parameters, the form of the potential (e.g. with or without short-range attractions), the spatial dimension, the pressure, and the system size. 
We can begin to see this result in Fig.~\ref{fig:rankJ}, where we show $\rank(J)$ as a function of $n_\theta$ for 2d and 3d systems with different combinations of relevant features. Here, we choose the parameters by assigning the particles to one of $n_\theta$ species types, with $2 \leq n_\theta \leq 40$, and setting $\bm \theta$ to be the diameters of these species. In each case, $\rank(J) = \min(n_y,n_\theta)$ exactly, verified through 10 independent measurements (different initial systems) for each data point. The inset scales both $\rank(J)$ and $n_\theta$ by $n_y$, more clearly showing that these systems are fully independent. In contrast, the triangles show similar data for an ordered system, with $\rank(J)$ plateauing well before $n_y$.

Figure~\ref{fig:singular_values} shows the susceptibilities (singular values of $J$) for 40 different scenarios that differ in features, parameters, potentials, spatial dimension, etc. For crystals (gray background), some if not most of the susceptibilities are numerically zero, meaning the features are only partially independent. For disordered solids, however, the susceptibilities are all clearly positive, demonstrating the robustness of our central result that mechanical properties are fully independent.
Of particular note, Scenarios 1-3 show no discernible change in the susceptibilities as the system size increases from $N=256$ to $N=4096$. A more quantitative finite-size analysis in the SI shows that the smallest susceptibility is roughly constant in $N$ and does not vanish in the thermodynamic limit. Interestingly, even though measures of anisotropic elasticity vanish for large $N$, the ability to tune in these directions remains finite. 

Central to the idea of independent response is its implications for tuning the features: independence \emph{means} independently tuneable, at least to linear order. 
The susceptibility measures how sensitive the features are to parameter changes, and we can see from Fig.~\ref{fig:singular_values} that this can vary by orders of magnitude.
Therefore, while the mechanical properties of disordered solids are fully independent, some combinations of properties are easier to tune than others.

\begin{figure}
    \centering
    \includegraphics[width=1.0\linewidth]{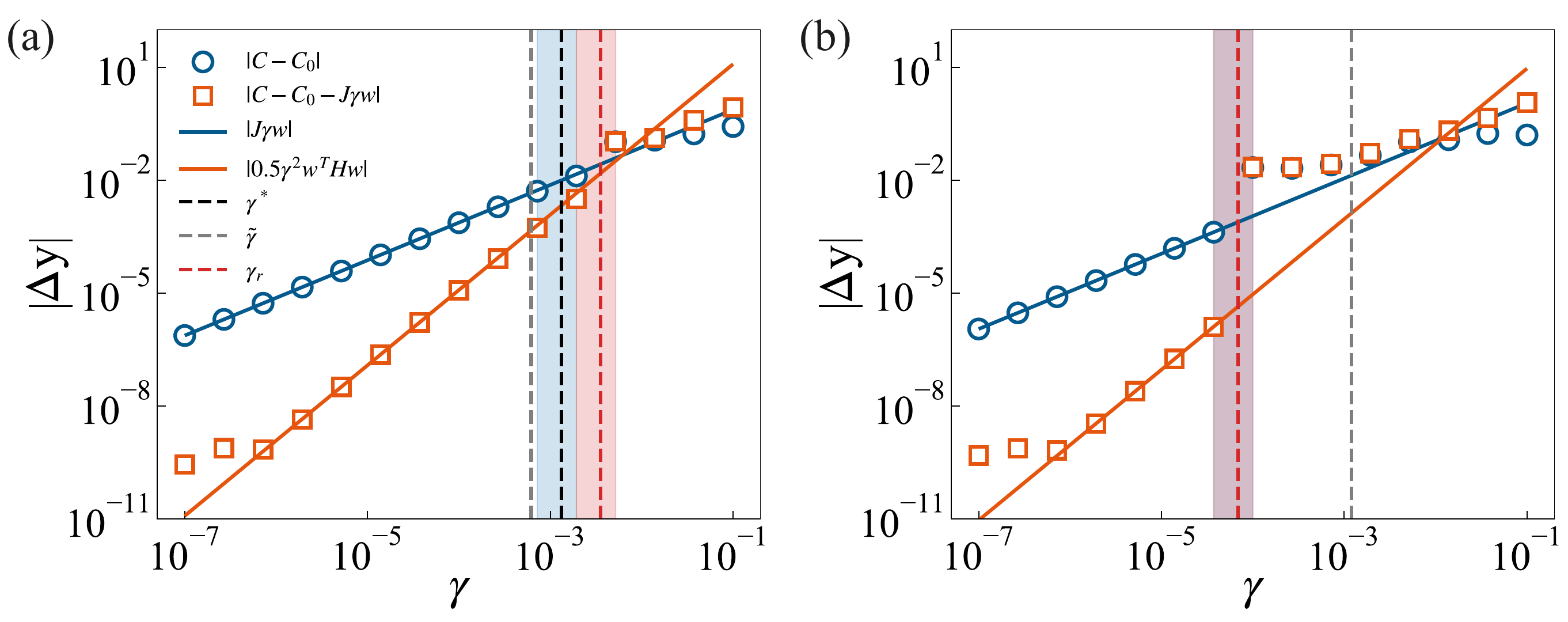}
    \caption{
    The extent of the linear regime. Measured feature changes (blue points) resulting from changing parameters by an amount $\gamma$ in a direction $\dth$ shows excellent agreement with the linear prediction (blue line) for small $\gamma$. Data is shown in a) and b) for two different example systems from Scenario 1 with $\dth = w_{0},w_{1}$.
    The difference (orange points) agrees with the $\gamma^2$ term in the Taylor expansion (orange line, see the Supplementary Material~\cite{Note1}). 
    The end of the linear regime ($\gamma^*$, black vertical line) is caused either by a) the increasing $\gamma^2$ term ($\tilde \gamma$, grey vertical line), or b) structural rearrangements ($\gamma_r$, red vertical line). The shaded blue and red regions denote the uncertainty in our measurements of $\gamma^*$ and $\gamma_r$, see the Supplementary Material~\cite{Note1}.
    }
    \label{fig:gamma_star}
\end{figure}

For small target changes to the features, $\dy^*$, we can solve the inverse problem in a single step:
\begin{align} \label{eq:Jinv_dy}
    \dt^* \approx J^+ \dy^*,
\end{align}
where $\dt^*$ is the change in parameters that results in the desired change in features, and $J^+$ is the pseudo-inverse of the Jacobian. 
What is the range over which this linear approximation is accurate? If we write parameter changes as $\dt = \dth \gamma$, with $\gamma \equiv \left| \dt \right|$, what is the maximum magnitude, $\gamma^*$, for which Eqs.~\eqref{eq:linear_approx} and \eqref{eq:Jinv_dy} are valid?
The blue points in Fig.~\ref{fig:gamma_star}a show $\left| \dy \right|$ as a function of $\gamma$, and the blue line shows the linear prediction $J\dth\gamma$. The difference (orange points) can also be predicted (orange line) from the $\gamma^2$ term in the Taylor expansion (see SI for details). We define $\tilde \gamma$ (shown by the vertical dashed gray line) to be when the $\gamma^2$ term becomes 10\% of the linear term. The solid vertical line shows our best estimate of $\gamma^*$ determined from where the actual $\dy$ deviates from $J\dt$ by more than 10\%. 

\begin{figure}
    \centering
    \includegraphics[width=1.0\linewidth]{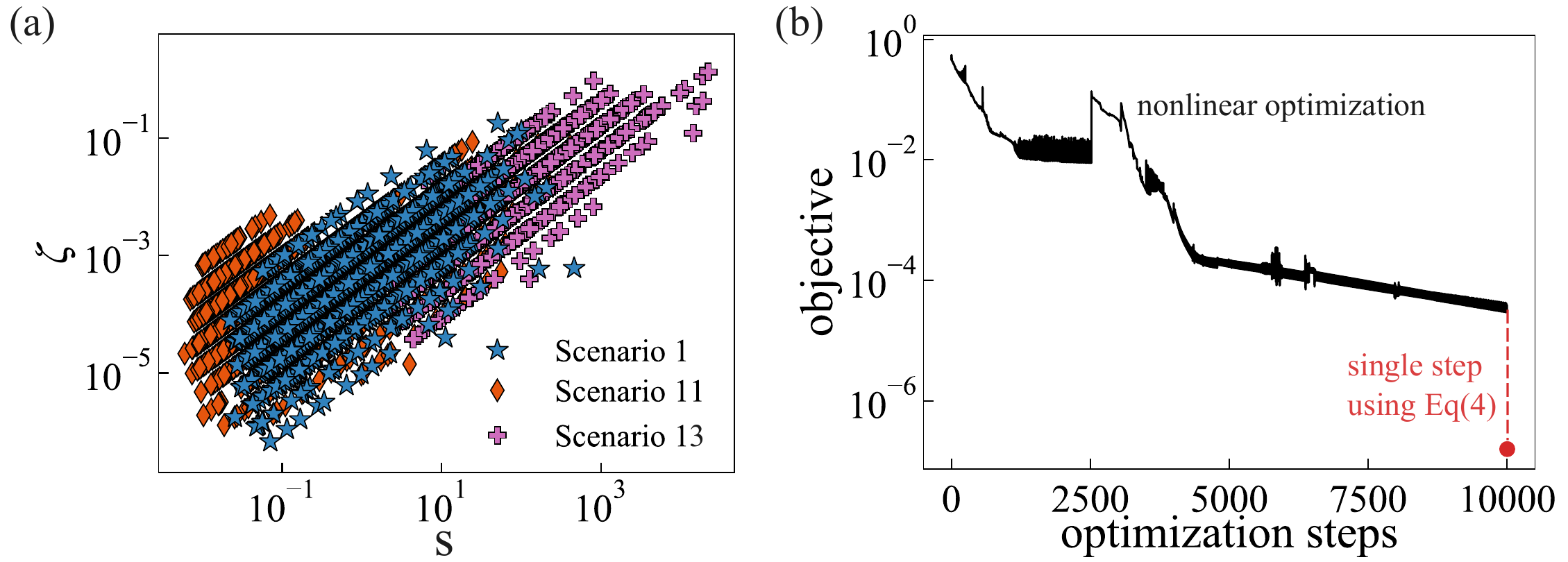}
    \caption{
    Maximum linear tunability and nonlinear optimization.
    a) The maximum linear tunability, $\zeta \equiv \gamma^* s$, is strongly correlated with $s$, meaning that feature combinations with high susceptibility have a greater potential for inverse design within the linear regime. 
    b) The objective $\mathcal{L}$ during nonlinear optimization of all six elastic constants in a 2d configuration from Scenario 15. Not only are the elastic constants fully independent, they can be tuned independently to predetermined targets well outside the linear regime.
    }
    \label{fig:design}
\end{figure}

In Fig.~\ref{fig:gamma_star}a, $\tilde \gamma$ provides an excellent prediction of $\gamma^*$ using only gradient information from the initial system. However, we find that structural rearrangements caused by parameter changes can take a system out of the linear regime well before $\tilde \gamma$. An example of this is shown in Fig.~\ref{fig:gamma_star}b. The vertical red line shows when the first structural rearrangement occurs, which corresponds to $\gamma^*$. While rearrangements become more frequent for large systems, the arguments of Refs.~\cite{Goodrich2014comment, Goodrich2014}, combined with the positivity of the smallest singular value as $N\to \infty$ discussed above, suggest that there is nevertheless a well-defined linear regime in the thermodynamic limit because the effect of a single rearrangement vanishes. 

%Figure~\ref{fig:design}a shows 
We find that the distribution of $\gamma^*$ in our systems is quite broad and largely uncorrelated with the susceptibility in the direction probed (Fig~S5 in the Supplementary Material~\cite{Note1}). Predicting $\gamma^*$ from the initial system is challenging because predicting rearrangements precisely is notoriously difficult. Nevertheless, we can use our measurements of $\gamma^*$ in directions of relevant design modes with susceptibility $s$ to define the ``maximum linear tunability'' $\zeta \equiv \gamma^* s$. This can be thought of as $\gamma^*$ in units of features, rather than parameters, and quantifies how far the features can be precisely tuned without resorting to nonlinear optimization techniques. As $\gamma^*$ is at best weakly correlated with $s$, Fig.~\ref{fig:design}a shows that $\zeta$ is correlated with susceptibility. Therefore, the susceptibility should be taken seriously as an indicator of the degree to which the features can be easily tuned.

Finally, we demonstrate that the ability to simultaneously tune the mechanical properties of disordered solids extends beyond the linear regime. In Fig.~\ref{fig:design}b, we simultaneously tune all six elastic constants of a 2d system from Scenario 15, with target elastic constants that differ by 10\% from their initial values. Following the numerical procedure outlined in Ref.~\cite{Zu2024}, we construct an objective function $\mathcal{L}$ that vanishes when all elastic constants reach the desired value, and use AD to calculate the gradient $\nabla_\theta \mathcal{L}$, enabling iterative nonlinear optimization (see SI). Over the course of 10000 optimization steps, $\mathcal{L}$ decreases by four orders of magnitude despite the system undergoing a structural rearrangement near step 2500. This gets the system close enough to the target behavior that the linear approximation is highly accurate, and we can lower $\mathcal{L}$ by an additional 2 orders of magnitude by a single application of Eq.~\eqref{eq:Jinv_dy}. Thus, all six constants have been simultaneously tuned to their target value, consistent with full independence. 

We have shown that, in a wide range of different scenarios, the mechanical properties of disordered solids are fully independent with respect to species-level parameters, while the same properties of crystals are not. This means that $n_y$ such features can be adjusted independently as long as there are at least as many adjustable control knobs $\bm{\theta}$: $n_\theta \geq n_y$. 
%While other work has shown high levels of independence beyond that of crystals, 
Independence is defined from the singular values of the Jacobian, $J$, obtained by expanding to linear order the change in relevant features, {\it e.g.} elastic constants, that results from changing relevant parameters, {\it e.g.} species' diameters. 
The rank of $J$ %({\it i.e.} the number of positive singular values) 
is the number of ways the features can be changed independently, and the magnitude of the singular values quantifies the susceptibility. This susceptibility can change by orders of magnitude, meaning that some feature combinations are more easily tuned than others, and is correlated with the maximum linear tunability. Furthermore, features can be tuned simultaneously and precisely well beyond the linear regime.
These results quantify, generalize, and solidify our understanding of a fundamental difference between order and disorder~\cite{Kittel2005, AM1976,Zaccone2023, Xu2011,Song2008,Zamponi2010,Zeller1971,Pohl2002,ramos2022low,Tsuneyoshi2002,Giancarlo2001,Xu2007,Cubuk2017}, and provide a mathematical foundation for the inverse design of materials for desired properties. 

This formulation also sets the stage for further understanding of tunability in specific scenarios. For example, Scenario 12 in Fig.~\ref{fig:singular_values} shows a band of small but non-zero susceptibilities. Inspection of the corresponding design modes shows that the large susceptibilities are strongly associated with diameter changes while the small susceptibilities are associated with changing the energy scales. This result is consistent with previous observations that tuning energy scales are less impactful than tuning particle diameters~\cite{Hagh2022}. Furthermore, this low-susceptibility band scales with pressure, see Fig.~S3 in the SI.

Moreover, our analysis raises a number of interesting questions. First, what is the nature of the transition from ordered to disordered? It is known that very small amounts of disorder can dominate mechanical properties~\cite{GoodrichNatPhy2014}, suggesting that full independence may be attainable in polycrystals or highly defected crystals. Next, to what extent do susceptibilities inform non-linear inverse design at the ensemble level, thus enabling design of experimental systems? Ref.~\cite{Zu2024} demonstrates simultaneous nonlinear optimization of the ensemble-average pressure and Poisson's ratio, but a thorough understanding of independence between many ensemble-averaged quantities would directly predict the potential for multifunctional design in experiments. 

Finally, while we have found full independence in all of our disordered scenarios, it is certainly possible that exceptions do exist. In fact, Rocks et al.~\cite{Rocks2019} find that the number of desired microscopic allosteric responses that can be tuned simultaneously is sublinear in the number of degrees of freedom. Is this predicted by incompatible feature strains? Can the Jacobian be used to find nonrandom allosteric features that display full independence? 

Finally, our definition of independent response necessarily depends on the parameterization. It is not correct to say that features are intrinsically independent, they can only be independent \emph{with respect to} the given parameters. These can be bond- or particle-level, species-level, or system-level parameters, speaking to the immense generality of this approach. Furthermore, systems do not need to be athermal particle packings, and the relevant features are not restricted to mechanical or elastic calculations. Rather, one can quantify the independence of any set of features with respect to any set of parameters in any system -- all that is needed in practice is the ability to accurately calculate the Jacobian. Our approach of employing Automatic Differentiation to efficiently and accurately calculate the Jacobian is, in principle, widely applicable with ever-growing differentiable-programming ecosystems~\cite{jax2018github,jaxmd2020}. Thus, we expect our results to inform inverse design in a wide range of settings, ranging from enhancing nonreciprocal elasticity to discovering time-dependent annealing protocols to the assembly of complex and functional nanomachines. 

\nocite{*}

\bibliography{manuscript2}% Produces the bibliography via BibTeX.

\end{document}